RESEARCH-ARTICLE

# EventChat: Implementation and user-centric evaluation of a large language model-driven conversational recommender system for exploring leisure events in an SME context

**HANNES KUNSTMANN**, Swiss Federal Institute of Technology, Zurich, Zurich, ZH, Switzerland

**JOSEPH OLLIER**, Swiss Federal Institute of Technology, Zurich, Zurich, ZH, Switzerland

**JOEL PERSSON**, Swiss Federal Institute of Technology, Zurich, Zurich, ZH, Switzerland

**FLORIAN VON WANGENHEIM**, Swiss Federal Institute of Technology, Zurich, Zurich, ZH, Switzerland

# EventChat: Implementation and User-Centric Evaluation of a Large Language Model-Driven Conversational Recommender System for Exploring Leisure Events in an SME Context

HANNES KUNSTMANN[*]

Chair of Technology Marketing, ETH Zurich, Zürich, Switzerland, h.kunstmann@outlook.com

JOSEPH OLLIER

Mobiliar Lab for Analytics, ETH Zurich, Zürich, Switzerland, jollier@ethz.ch

JOEL PERSSON

Chair of Technology Marketing, ETH Zurich, Zürich, Switzerland, joel.persson.91@gmail.com

FLORIAN VON WANGENHEIM

Chair of Technology Marketing, ETH Zurich, Zürich, Switzerland, fwangenheim@ethz.ch

The integration of large language models (LLMs) to conversational recommender systems (CRS) represents an enormous evolution in their strategic potential. Yet to date, research has predominantly focused upon technical frameworks to implement LLM-driven CRS, at the expense of end-user evaluations or strategic implications for firms, particularly from the perspective of a small to medium enterprises (SME) that makeup the bedrock of the global economy. In the current paper, we detail the design and field performance of an LLM-driven CRS in a small to medium enterprise (SME) context using both system metrics and end-user evaluations, while also presenting a revised ResQue model for evaluating LLM-driven CRS, enabling replicability in a rapidly evolving field. Results demonstrate satisfactory system performance (85.5% perceived recommendation accuracy) but underscore latency, cost, and quality challenges. Notably, with median costs of $0.04 per interaction and latency of 5.7s, cost-effectiveness and response time emerge as crucial issues, predominantly driven by use of ChatGPT as a ranker within the retrieval-augmented generation (RAG) technique. Results also show that relying solely on prompt-based learning has quality limitations in a production environment. Strategic considerations for SMEs are outlined considering trade-offs in the technical landscape.

* These authors contributed equally and share first authorship
† Corresponding author

## 1 INTRODUCTION

The importance of recommender systems (RS) to simplify user decision-making and cope with information overload has been well-acknowledged in both information systems research [24, 100, 103] and practitioners circles alike [12, 84]. Recommendations can account for as much as 30% of firm revenue, with a 1% improvement in recommendation quality translating into billions of dollars [94]. A key limitation of traditional RS, however, is lack of user control over the recommendation process [6, 16, 53, 59, 60], with users constrained to reactively make choices among recommendations pre-integrated into the system's logic [3] rather than proactively describing their desired choice set [52]. Conversational recommender systems (CRS) overcome these challenges by explicitly allowing for both user input and feedback on suggestions (e.g., "This product is too expensive") with a conversational interface (i.e., a chatbot) supplemented by machine learning (ML) techniques [18] that refine recommendations and empower users in their search [50, 52].

The emergence and integration of large language models (LLMs) "capable of understanding and generating natural language and other types of content to perform a wide range of tasks" [46] to create LLM-driven CRS, a form of Gen-AI enabled RS (otherwise known as Gen-RecSys [23]), therefore represents a monumental shift in the strategic potential of RS for firms [79]. Beyond LLMs well-known competences in natural language processing and user engagement [8], their versatility makes them also exceptionally well-suited for architectural tasks such as serving as the RS [27] or reranking within the retrieval-augmented generation (RAG) technique [32]. LLMs can be seamlessly integrated into CRS architecture without requiring extensive data or other costly resources for model training, with even the most advanced LLMs accessible via simply calling application programming interfaces (APIs). LLM-driven CRS therefore represent a high-potential tool to create efficacious CRS systems, as recognized by leading industry players [45, 93]. However, for small-to-medium size enterprises (SMEs), that often face resource constraints or fine-margins, the path to implementing such strategically valuable systems remains unclear.

To date, several relevant frameworks for building LLM-driven CRS have been proposed, addressing practical challenges such as effective conversation management or the extraction of information from external sources [29, 32, 36, 44]. While greatly advancing the Gen-RecSys field, existing frameworks have predominantly outlined state-of-the-art methods in controlled settings, rather than addressing the constraints faced in real-world deployment. Specifically, to date, there remains a lack of research on the technical considerations faced when implementing an LLM-driven CRS for a given business context, and an evaluation of how valuable these implementations may be for SMEs from a strategic and user-experience perspective. Business-critical factors that remain ambiguous in existing frameworks include development costs due to technical complexity, operational expenses, as well as performance and latency implications such as whether to adopt an agent-like or stage-like architecture and how to contextualize LLMs for generating user responses [29, 32, 36, 44]. As a novel technology, uncertainty regarding system performance in production environments from an end user perspective also exists, raising further questions about the practicality and effectiveness of these systems for SMEs [116]. Additionally, for resource-constrained SMEs, the value of investing in additional features to further anthropomorphize the CRS or augment the UI (e.g., using clickable buttons) as widely recommended in extant chatbot and CRS research [41, 54, 92, 115] remains unclear.

In this paper, we investigate this phenomenon by taking the perspective of an SME in the leisure industry, developing and implementing a LLM-driven CRS and subsequently validating its performance the field. In the first part of the paper, we detail the design of a ChatGPT-driven CRS, delving into practical design choices, respective strengths and limitations, and focusing on business considerations specifically tailored for an SME context. Our first contribution is to demonstrate how various components can be brought together in a resource-efficient system that effectively interacts with real customers in the field, rather than detail the state-of-the-art development of a single component individually. In the second part of the paper, we evaluate system performance from an end-user perspective, using both objective (i.e., system-based user interaction metrics) and subjective (i.e., user-evaluations) measures to create a holistic picture of the system performance. As such, our second contribution is to provide an updated theoretical framework for the Gen-RecSys community, thus that user evaluations of LLM-driven CRS can be anchored within a common conceptual model. Creating a shared approach to user evaluations, in a field where new system components and foundational models are released every few months, will both facilitate replicability in the short-term and theoretical consensus in the long-term. To achieve this, we present a revised, short-form ResQue model [105] rooted in extant CRS literature and updated for use with LLM-driven CRS in the field. Lastly, in the third part of the paper, we discuss technical, strategic, and theoretical implications surrounding the utilization of LLM-driven CRS in SME contexts. This includes a discussion on the strengths and limitations of the current system design, and outlook for their design considering newer LLMs and other Gen-RecSys developed to date. Consequentially, we contribute to the current discourse on the applicability and challenges facing the deployment of LLM-driven systems in business settings, and more specifically, on the feasibility and usefulness of such systems for SMEs.

In sum, we:

1. Present *EventChat*, a resource-conscious LLM-driven CRS deployed by an SME.
2. Provide field evaluation with both subjective (ResQue) and objective (token and latency-level) metrics.
3. Extend and validate a revised ResQue model for LLM-driven CRS.
4. Discuss SME-specific managerial trade-offs in system design and operation.

Taken together, our research aims to democratize the roll out of LLM-driven CRS to SMEs, which make up the bedrock of the global economy [113], whilst explicitly accounting for the costs and complexity that is critical for success in an SME setting [83].

## 2 RELATED WORK

### 2.1 Strategic Potential of LLM-driven CRS for SMEs

As an emerging technology, there exists little research that has empirically verified the strategic effectiveness of LLM-driven CRS in the field, particularly from an SME perspective [49]. The long-standing importance of RS, CRS, chatbots, and the recommendations they provide has been well established across a variety of industries however [75, 81]. Through their ability to relay information in a dyadic, low-effort manner, RS technologies can enhance consumer product search (i.e., greater exploration of product options), upsell (i.e., encourage purchases of higher monetary value), or dynamically adjust prices to consumer input (i.e., charge higher prices), improving core business outcomes such as volume or value of sales [72, 101]. Beyond their effects on sales alone, however, conversational forms of RS (e.g., CRS, chatbots) can facilitate increased user engagement, loyalty, and customer retention via the preference elicitation process [24, 51], closely mirroring the interpersonal dynamics that add value in human-based service interactions [7, 20, 97].

The advent of LLM technology and its implementation into CRS systems therefore represents a paradigm shift in the strategic potential of CRS [79]. By improving both the recommendations they provide [89], as well as offering a more natural and engaging conversational experience to users [122], LLM-driven technologies in customer facing roles have been estimated to increase productivity between 30-45% whilst also enhancing the customer experience [85]. Considering that for SMEs, fine-margins and relational outcomes dominate strategical positioning [121], and that humanized and personalized feedback are features of chatbots previously found to most readily distinguish an SMEs service offering compared to rivals [112, 114], the opportunity to deploy LLM-driven CRS in customer facing settings represents a high-value method to realize business value.

Nevertheless, for resource-constrained SMEs, leveraging novel technologies often proves challenging [113] with missed opportunities related to generative AI a present danger [1], possessing the potential to further exacerbate structural differences in the economy between smaller and larger firms [83]. Indeed, early evidence exists that "knowledge spillovers" usually enabling smaller firms to catch up or "leap frog" larger players are less available when it comes to AI technologies [69] due to lack of know-how or skilled personnel [104]. Even when such tools are adopted, the application of AI or LLM tools by SMEs does not necessarily guarantee success, as such tools must be tailored to the unique business and strategic context, with off-the-shelf solutions less likely to deliver the contextual, real-time personalization that AI can deliver [116]. There exists a clear need, therefore, to understand how to develop a LLM-driven CRS to the specific business circumstances of an SME. By doing so, SMEs can increase their efficiency in customer interactions, offset the resource constraints they face [113], and deliver the personalization critical of commercial success in today's data-driven markets [10, 61, 91].

### 2.2 Frameworks for LLM-driven CRS

While the opportunity presented by LLM-driven CRS for SMEs is clear, the rapid scaling-up of data and parameters has made the route to implementation less so. Enhanced capabilities such as in-context learning (ICL) [14, 78, 110], adherence to instructions [120], and planning and reasoning [125, 127, 128] have the potential to create state-of-the-art CRS. However, limitations regarding complex user inquiries and maintaining seamless communication with users remain for such systems. For SMEs, this is a significant barrier due to the relative weight each customer interaction has in overall business performance, and the reality that SMEs are less able to absorb the effects of occasionally poor customer service or change to alternative system designs when compared to larger rivals [35].

To overcome these customer experience challenges and enhance the overall user experience, three prominent frameworks have been introduced that use LLMs in CRS [29, 32, 44]. These aim to: (i) better understand and control conversations effectively, (ii) retrieve information from external sources, (iii) deal with insufficient conversational data for training, (iv) address potential limitations of LLMs in capturing fine-grained, domain-specific behavior patterns, and (v) manage the unpredictability and proneness to hallucinations in LLMs. Table 1 provides an overview of these frameworks.

Table 1: Overview of frameworks for LLMs in CRS.

| Framework | Use of LLM in CRS | LLM Used | Function |
|---|---|---|---|
| RecLLM [32] | As a dialog management module for orchestrating calls to other modules | Fine-tuned version of LaMDA LLM [118]. | To rank slate recommendations from an RS so as to match user preferences and enhance personalization by adjusting session contexts to natural language-based user profiles |
| InteRecAgent [44] | As the logic for discerning users' intentions | GPT-4 | To trigger sequences of API calls for generating conversational responses based on outcomes from RS tools |
| LLMCRS [29] | As an orchestrator of subtasks | Flan-T5-Large [21] and LLaMA-7b [120] | To divide the RS workflow into stages of subtasks that are performed every iteration |

The first framework, RecLLM [32] is a LLM-driven dialog management module that orchestrates calls to other modules. It uses a fine-tuned version of LaMDA as the underlying LLM [118]. RecLLM uses a recommendation engine to retrieve items and introduces an LLM-driven ranker module to match preferences with item metadata and generate a slate of recommendations to display to the user. The framework also applies interpretable user profiles in natural language to adjust session-level context and enhance personalization. Second, the underlying idea of the InteRecAgent (Interactive Recommender Agent) framework [44] involves using an LLM (specifically the GPT-4 model) as the brain and uses RS as tools. The LLM discerns users' intentions and assesses whether the ongoing conversation requires using these tools. If so, the LLM triggers a sequence of API calls to generate a response based on the outcomes of the execution of the tools. Third, the framework LLMCRS [29] focuses on the orchestration of sub-tasks within a CRS. LLMCRS divides the workflow into stages of subtasks that are performed every iteration. The input processing and the output generation are performed by the LLMs Flan-T5-Large [21] and LLaMA-7b [120] respectively. To adapt these foundational LLMs to conversational recommendations, they propose fine-tuning the LLMs using reinforcement learning from CRS performance feedback.

To date, it remains unclear which of these three frameworks provides methods best suited to a specific business context, with this ambiguity representing a major hurdle for enterprises aiming to introduce LLM-driven CRS. While this issue is not unique to LLM-driven CRS contexts alone [39], it is arguably more critical to address them in LLM-driven CRS as the associated costs of integrating LLMs are considerable, particularly for SMEs. Understanding how relevant aspects of each framework can be utilized, whilst also paying heed to unique contextual business and strategic factors when implementing a LLM-driven CRS remains an open question. To fill this gap, we move beyond existing frameworks by detailing a holistic, end-to-end LLM-driven CRS designed specifically for SME event recommendation, named EventChat. While prior systems emphasize fine-tuning (RecLLM), agentic tool orchestration (InteRecAgent), or staged subtasks with fine-tuned models (LLMCRS), EventChat contributes a pragmatic design optimized for the resource-constraints faced by SMEs. Specifically, by introducing: (i) a hybrid chat + UI where the visibility detector feeds viewed items back into the conversational context, (ii) a fixed five-action controller embedded in a stage-based pipeline to ensure stability and cost-efficiency, (iii) slate-level re-ranking that filters candidate slates in a single prompt to reduce computational cost, (iv) digest-based item inquiry, which uses structured item digests in place of tool-based SQL generation, and (v) pure ICL eliminating the need for training data, which is hard to obtain in practice, and costly fine tuning. These architectural decisions allow the system to remain deployable under real SME resource limitations, while still covering search, recommendation, and targeted inquiry to deliver a service relevant to real users in a real business context.

### 2.3 Evaluation of LLM-applications

In addition to understanding technical choices when implementing a LLM-driven CRS, of equal importance are customer ratings of the system and its economic viability once implemented. While extant research has examined LLM-driven applications across diverse fields, including finance [129], education [64], chemical engineering [73], as well as domain-independent frameworks [55, 119], these works have concentrated on evaluating performance against state-of-the-art techniques rather than end user evaluations. Indeed, a recent review by Manzoor and Jannach [82] found that the majority of current CRS research relies on comparative evaluation or human annotators rather than real user feedback and business-critical factors such as system costs and latency from field applications.

In traditional RS domains, the well-established ResQue framework has been applied to evaluate users' subjective ratings of RS usage experiences [105]. With origins in cognitive-affective behavioral models applied in information systems research (e.g., Technology Acceptance Model, Expectation-Confirmation Theory), the ResQue model states that user-perceived quality factors (e.g., recommendation accuracy) influence users' beliefs about the system (e.g., perceived usefulness), in turn shaping user attitudes (e.g., satisfaction), which lead to behavioral intentions (e.g., intention to use the system) [105]. A core strength of the framework has been its ability to provide a unified system to evaluate aspects of the usage experience across diverse application settings such as e-commerce [105], music [57], travel [11], and movies [102]. By doing so, it reflects the value attributed to the holistic usage experience of the RS within its application domain, rather than simply measuring the performance of the underlying recommendation algorithm [105], moving beyond a focus on task efficiency alone to wider factors that sustain user engagement, for example, transparency and controllability [105], or perceived novelty and diversity of recommendations [106].

The applicability of ResQue to CRS and LLM-driven CRS has been questioned in recent years on two counts however: First, as RS have evolved into interactive, dialogue-based CRS powered by AI [49] the original ResQue model is unlikely to fully reflect the current user experience. For a LLM-driven CRS, for example, factors relating to Interaction Quality (e.g., Consistency, Coherence, and Input Processing Performance) have potentially far greater relevance than traditional RS systems with rigid (e.g., script-based) conversational flows. Indeed, recent applications of communication theories to human-AI interactions, for example Grounding Theory [22] have suggested that utterance clarity is key to facilitating a shared mental model of intended behaviors and outcomes [25]. However, to date, factors related to conversational clarity are entirely absent ResQue models applied in practice, thus highlighting the need for a revised ResQue for LLM-driven CRS.

Second, to compliment such subjective evaluations, there is a pressing need to examine users' experiences of the LLM-driven CRS alongside objective user-interaction data. Doing so could anchor subjective user evaluations in real system events from the actual usage context, giving specific information about why a certain construct (e.g., Perceived Usefulness) scored poorly [49]. While the importance of combining objective results with subjective feedback has been strongly recommended within the RS field, for example by Knijnenburg et al. [67] who argued that subjective evaluations are essential in illuminating contradictions in system performance or Ge et al. [38] who incorporated novel construct alongside behavioral data to uncover user tradeoffs, this task has not yet been completed for Gen-RecSys. While recent research has suggested revising ResQue to capture the more complex conversational dynamics presented in human-AI interactions [49, 58], empirical validation with end-users using a LLM-driven CRS has not yet occurred. There exists therefore a clear need to extend and test a revised ResQue model to LLM-driven CRS settings, with adaptations that address the interactive, conversational characteristics of CRS, whilst also anchoring these ratings in real system events (i.e., objective system metrics). Such a multi-method approach would both conform with recommendations to update ResQue with the advent of AI-technologies [49], as well as with recent guidelines for making information systems research more relevant to practitioners [88].

### 2.4 Positioning EventChat within the Gen-RecSys landscape

While previous frameworks collectively advance the technical sophistication of Gen-RecSyS via approaches such as fine-tuning, agentic orchestration, or staged subtasks, they remain primarily oriented toward demonstrating state-of-the-art performance under controlled conditions. In contrast, the present work emphasizes deployability under the constraints faced by SMEs, whilst also collecting feedback from real users in the field, to verify the systems effectiveness from a user perspective. EventChat thus contributes not by new algorithmic components, but by integrating existing techniques into a resource-conscious end-to-end system that can be operated with limited engineering capacity. This perspective complements prior frameworks by illustrating how LLM-driven CRS can be adapted to real business settings where costs, latency, and maintainability are decisive.

## 3 BUSINESS CONTEXT

To address these design considerations and understand their impact on the usage experience from an SME perspective, we partnered with a startup in the leisure industry which aimed to adopt AI technology to enhance leisure events exploration and planning. For this purpose, we collaboratively implemented EventChat, a LLM-driven CRS introduced as a new feature in the latest update of the startup's existing iOS and Android applications. The system complemented the already available methods for users to explore events and activities in a major German city, such as a filter-based search, an interactive map, and a calendar overview.

### 3.1 Business Motivation for EventChat

EventChat was designed to offer general recommendations, empower users with search functionality, refine the recommended slates based on user feedback, and inquire the system about details of events. These features were motivated by two key objectives of the startup: (i) to elevate the user experience through improved leisure option discovery, and (ii) to address and mitigate potential data quality issues arising from the startup's web scraping methodology.

### 3.2 Enhanced User Experience

Through this project, the startup aimed to improve the user experience on its platform. Being a transactional platform for events and activities, the seamless and user-friendly acquisition of information was at the core of its value proposition. The startup referred to events as entertainment opportunities with defined start and end dates, such as concerts or theatre performances, while activities were organized by their operating hours, encompassing venues like public pools, bars, and restaurants. By providing information about both events and activities, the startup positioned itself as a digital platform that allowed users to inform themselves about an entire range of leisure opportunities. This extensive selection of leisure choices, however, resulted in over 81 (sub-) categories. One objective for this project was therefore to enhance the exchange of information on the platform by providing a novel and interactive functionality to explore leisure opportunities. Note that in the following we refer to events and activities just with events for simplicity.

### 3.3 Mitigation of Web-scraping Side-effects

The startup's platform sourced its event listings through two methods: direct submissions from event organizers and web scraping using the startup's algorithms applied across various websites. Direct submission gave event organizers agency to be discovered, in particular for less well-known events, whereas web scraping was integral for extensive coverage. Together, the two methods ensured a diverse range of event information was available. However, the web scraping method presented challenges in maintaining data quality. For instance, depending on the structure of the source website, essential details like price information might not have been captured by the scraping algorithms. As a result, important information embedded in the event descriptions on the web-scraped source website may have been absent from the relevant fields of the startup's relational database. In contrast to a common filter-based search, however, LLM-driven CRS can be designed to parse and interpret the details of the event descriptions, mitigating the unwanted side effects of web scraping.

### 3.4 Synthesis for Business Context

The dual objectives of elevating user experience and mitigating data quality issues created a particularly demanding context for the implementation of a CRS. For the startup, the challenge was not only to support efficient leisure option discovery in a complex and heterogeneous space, but also to compensate for structural imperfections in the underlying data. This combination of user-facing and firm-facing constraints makes the SME setting especially suitable for studying the practical feasibility of LLM-driven CRS, as it highlights both the opportunities of conversational interaction and the engineering and deployment trade-offs that SMEs must navigate.

## 4 SYSTEM DESIGN

### 4.1 Business-driven Design Choices

In choosing the technical design of EventChat, we deliberately prioritized feasibility and appropriateness for SMEs over state-of-the-art algorithmic sophistication. The resulting design was therefore less about optimizing a single component in isolation, and more about demonstrating how such systems can be deployed under the resource and skill constraints of a real SME. Our first considering was the underlying LLM used by the CRS. ChatGPT utilizes many of the novel LLM capabilities including in-context learning [14, 78, 110], adherence to instructions [120], as well as planning and reasoning [125, 127, 128]. When we developed EventChat in August 2023, ChatGPT was the most advanced LLM available on Azure OpenAI Service. The startup, through its membership in the Microsoft for Startup Founders Hub, used program credits for access and thereby avoided direct costs for calling OpenAI's ChatGPT API. For this reason, ChatGPT was selected.

Second, we implemented the CRS relying on prompt-based learning. Our approach drew inspiration from Huang et al. [44] who used GPT4 as the underlying LLM. Our choice of prompt-based learning was driven by its numerous benefits for SMEs over alternative approaches like fine-tuning. In particular, prompt-based learning does not require large amounts of training data from past user interactions or generated synthetic data, both of which are costly to obtain for SMEs in terms of time, finances, and engineering efforts. Additionally, the frameworks that necessitate fine-tuning rely on base models with significantly fewer parameters than ChatGPT [29, 32]. Hence, prompt-based learning offers a CRS implementation that suits the constraints faced by an SME.

Third, the cost implications associated with calling an LLM API necessitated a design that prioritizes task-oriented interactions over phatic (i.e., small-talk) communication. This economic consideration supported the implementation of an attribute-based question-answering CRS [65, 137, 138]. In attribute-based question-answering, the goal is to suggest appropriate items to users in as few rounds as possible. The system and users mainly engage in question-answering about desired item attributes, gradually adjusting user interests. Notably, this approach went against the dominant approach in extant chatbot research that emphasizes the value of increased anthropomorphization of chatbots [96].

Fourth, for recommending events to users, the time and location were crucial factors since users generally plan their leisure activities on a short- to mid-term basis and prefer events that are conveniently located near them. As the startup operated in in a single German city, we made no further location-wise adoptions within EventChat as the to-be-suggested events were already tailored to the local audience.

### 4.2 Architecture

In contrast to previous research [29, 32, 44], we defined EventChat not only as the back end but as the whole end-to-end system. Figure 1 provides a high-level overview of EventChat's architecture, which implemented a turn-based dialog system. The back end may have called endpoints of external resources like the startup's relational database, vector database, recommendation engine, or internet-based information sources.

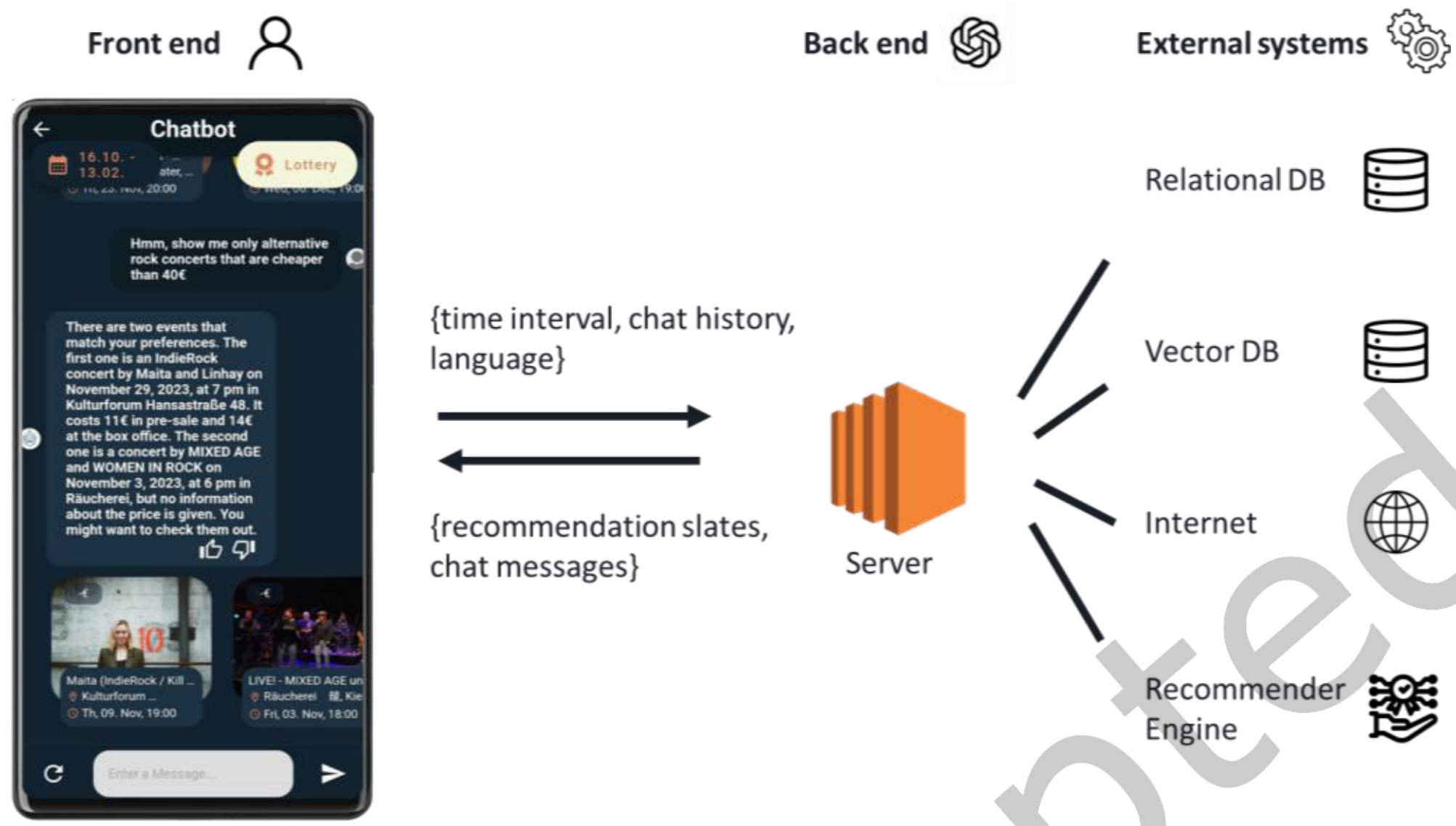

Figure 1: Architecture of EventChat

### 4.3 Front End

Given the startup's app-first strategy, we purposefully designed the front end for iOS and Android using the Flutter framework [31]. A distinguishing characteristic of event discovery is its inherent time dependency. To address this, we configured EventChat to consider events occurring within the next 150 days by default. Nevertheless, users could define a custom time interval using a button statically located in the top left corner of the interface. To enhance relevance, we included this user-specified period as a query parameter to EventChat.

Furthermore, the CRS implementation gave users the choice to either search for specific events or receive general recommendations based on their past preferences. In each session, users were prompted to make this selection by simply clicking buttons within the chat interface. Such a hybrid approach of including buttons in a chat interface has been found to increase user-perceived control [47, 92]. However, our main rationale behind including this feature was to optimize the usage of resources and economize on costs, as it helped us reduce iterations and calls to the LLM API.

Previous research has shown that users anticipate the presentation of information about items through a dedicated recommendation slate, which allows for easy access to additional details. [32, 54]. To contextualize the LLM around user interactions, we therefore used a visibility detector feature in the front end, which detects the last three event card summaries that were shown to the user. This information was then sent to the back end to provide the LLM with context. These features are shown in Figure 2.

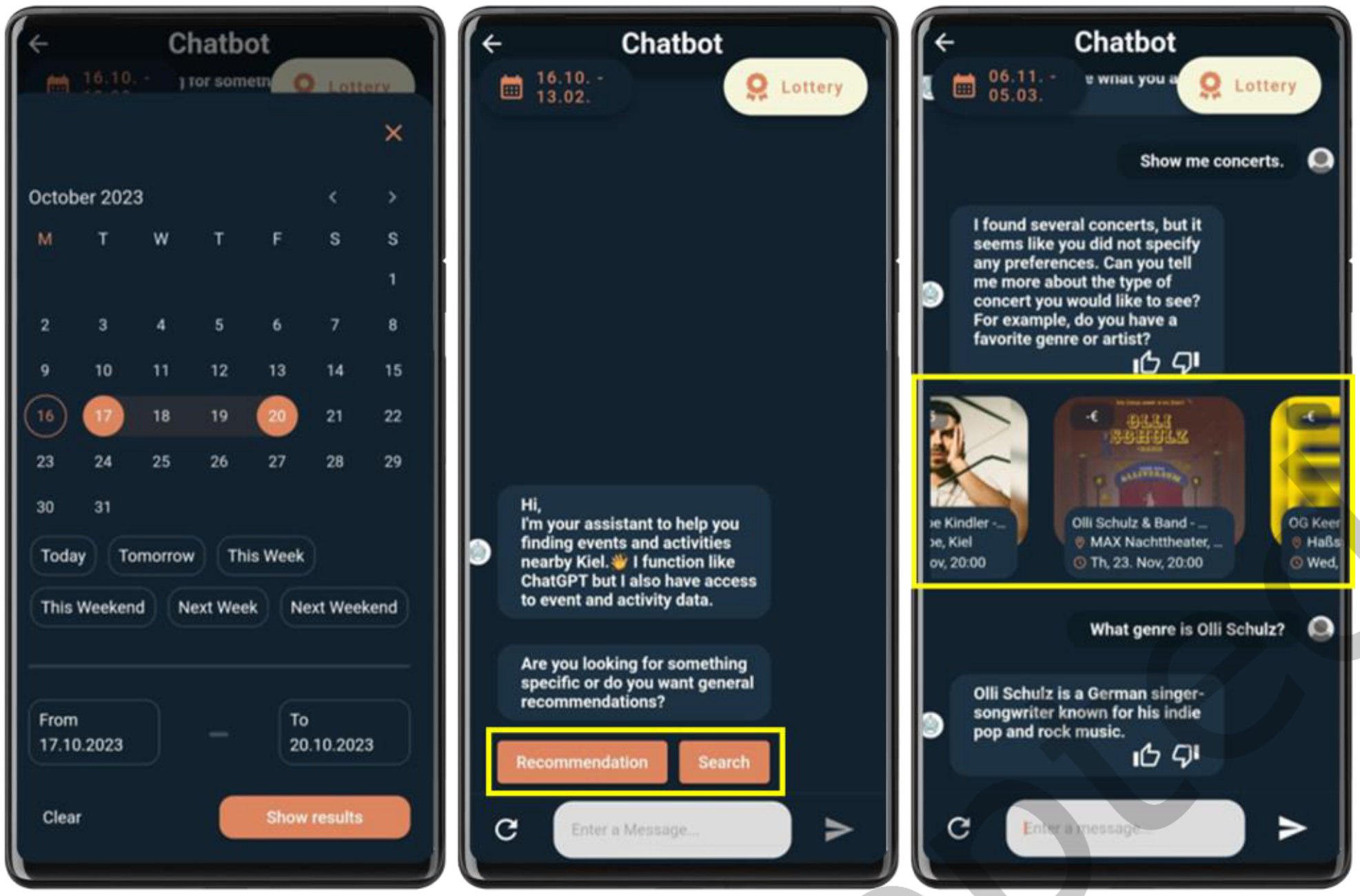

Figure 2: Illustration of time interval interface (left), case selection (middle), and highlighted visibility detection in recommendation slates (right) within EventChat's front end

## 4.4 Back End

### *4.4.1 Overview*

A conversation with EventChat was divided into turns, with each turn initiated by the user taking an action in the front end, such as answering in the chat interface. As a result of every turn, EventChat answered with a response to the user and, depending on the context, may have provided recommendation slates (Figure 3).

Our stage-based approach was inspired by Feng et al. [29, 32]. Like their framework, the workflow of EventChat was roughly divided into stages consisting of sub-tasks performed by an LLM in each iteration. We also tried an agent-based approach to enhance the flexibility of the system and thereby enable more sophisticated functions (e.g., whole-day trip planning). We soon, however, noticed limitations of this architecture in our context: (i) an agent architecture can lead to excessive LLM calls, which cause high costs and latency, as previously discussed by Wang et al. [126] and (ii) ChatGPT-specific performance issues recognizing and using the tools correctly as found in practitioner circles (see GitHub ticket by gcsun [37]). We therefore prioritized stability and limited latency as well as token usage over the flexibility an agent-based structure might have introduced.

At the beginning of every turn, the Global Context Storage was initialized. The stored context included the preferred language of the user, the time interval of interest, and the chat history. The chat history was stored in textual form to make it accessible to the LLM via ICL. Stages within the workflow could then retrieve or update information from it. Subsequently, one of the five following actions was chosen by the LLM: (i) Chat: EventChat directly answered the user based on the context; (ii) Refusal: EventChat responded with a pre-defined message since the user response was either inappropriate or off-topic; (iii) Search: EventChat initialized the search workflow, creating a recommendation slate of items based on the user's query derived from the current chat history; (iv) Recommendation: EventChat initialized the recommendation workflow which, creating a recommendation slate based on the user's preferences derived from past interactions with the startup's app, or (v) Targeted Inquiry: EventChat started the Targeted Inquiry workflow that aimed to answer a user's question for a specific event by gathering additional information via a database query or website.

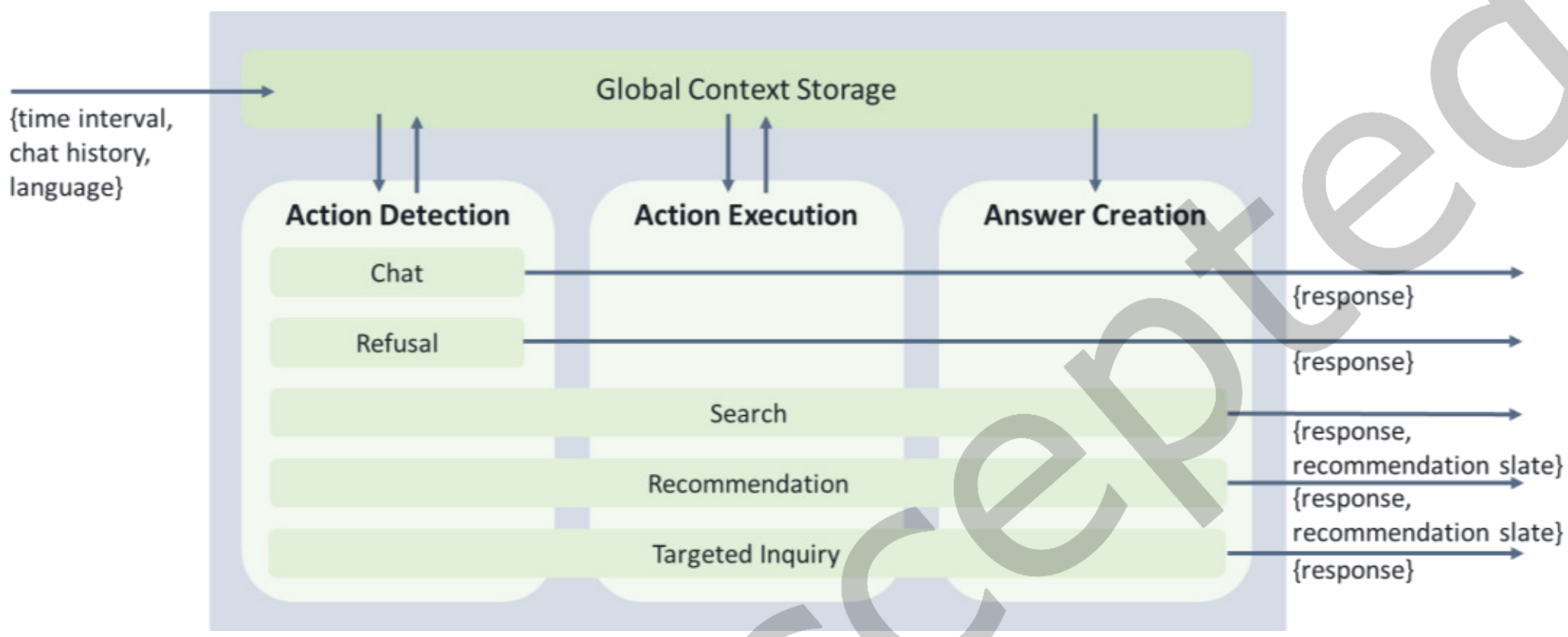

Figure 3: Conceptual architecture of EventChat's back end

### *4.4.2 Prompt Design*

The stages within EventChat's architecture used ChatGPT to perform dedicated tasks. When designing the corresponding prompts, we faced trade-offs regarding the quality of the LLM's response with latency and cost. More specifically, we noticed that the quality of the LLM's responses deteriorated as our prompts became more complex. In our design iterations, complexity was associated with prompts containing a greater amount of information, instructions, and tasks, which in turn resulted in longer overall prompt length. To enhance the quality of LLM responses, it thus would have been advantageous to break down each sub-task into separate prompts. However, this approach would have come at a higher cost: For each call, output format instructions would have needed to be repeated, resulting in more token usage and higher latency due to the increased round-trip time required for multiple calls. Furthermore, dividing tasks into several prompts would not always have been feasible as long prompts may be needed to provide the LLM with important contextual information regarding business specifics. After experimenting with several prompt versions, we eventually found solutions we considered optimal to balance these trade-offs. Our final prompts were designed to minimize the number of LLM API calls while still consistently delivering high-quality responses. Notably, in our back end, the boundaries of each of the stages were often driven by the extent of tasks that could be processed within the corresponding prompt. For instance, a response for the Chat action could directly be generated within the prompt of the Action Detection phase, making a distinct prompt for this action redundant.

From a prompt-design perspective, we employed techniques such as Few-Shot CoT and Few-Shot ICL to guide the LLM in performing business-specific tasks [68]. Additionally, we used schema-based format instructions to accurately interpret the LLM's outputs, following the methodology outlined in the langchain implementation [17].

#### *4.4.3 EventChat Action Modules*

We implemented three dedicated modules for the Search, Recommendation, and Targeted Inquiry actions. The Search and Recommendation workflows were fundamentally similar and show parallels to the RAG technique [32, 74]. First, EventChat generated a candidate set using the startup's existing retrieval infrastructure, either based on the user's query (Search) or past interactions (Recommendation), e.g., filtered retrieval via the recommendation engine or similarity-based retrieval when keywords could be extracted from user input. The selected time interval was forwarded as an explicit query parameter to constrain retrieval before candidates were provided to the LLM. To get a highly relevant candidate set ordered by relevance we applied filters to the startup's Amazon Personalize instance or the startup's vector store if keywords could be extracted from the user's query. Notably, rather than performing a vector store similarity search on the user's chat input, we extracted keywords. Candidate set sizes for Recommendation was fixed at 100 on AWS Personalize. For Search, these varied widely depending on the chosen category in combination with the time filter, which by default covers a rolling window of 150 days starting from the current date, but can be adjusted via the UI.

Second, we used ChatGPT as a ranker to determine whether the events shown matched the user's intention. To support reranking, EventChat constructed textual event digests prior to inference by combining structured database attributes with available descriptive content. These digests were supplied to the LLM as contextual input, avoiding the need for dynamically generated database queries and improving robustness given the heterogeneous event data. The resulting candidates were then passed to the LLM in a single reduction step. This batching allowed the model to compare items in context while limiting the number of API calls and associated latency. In contrast to previous work [32, 44], we simultaneously rank up to 10 items per prompt. This number ensured only the most relevant recommendations were recommended, and avoided potential information overload for users, a problem known in extant RS research [103]. Furthermore, our ranking only distinguished if the event matches the user's intention or not. Finally, we created a user response by subsuming an answer based on the user's request or interest and the suggested events.

The Targeted Inquiry allowed users to get more specific information about an event. To do this, we generated a comprehensive textual description of an event. This description was either created based on information in the startup's database or obtained from an associated event website. Ensuring the inclusion of all existing information within the token limit of 4096 tokens of ChatGPT was feasible due to the sparse metadata, representing a business-specific particularity. Consequently, the Targeted

Inquiry action relied on the LLM to respond to a user's query using this contextual information. This approach contrasted previous work [44] where the LLM created an entire SQL statement based on context about the database structure to answer user queries. When we experimentally allowed the LLM to create comprehensive SQL queries for the startup's relational database to increase the flexibility of the module it failed. This limitation arose from the complexity of the startup's relational database structure, which adhered to the Boyce-Codd normal form (BCNF) and often required multiple table joins to derive insights. Additionally, due to web scraping, the database contained columns that were often sparse. Contextualizing the LLM for such business specifics would have however further complicated the querying process.

### 4.5 Interim Discussion of the System Design

The previous sections showcased how business requirements influence design choices and the trade-offs businesses face when designing an LLM-driven application. We now examine these aspects further:

First, regarding the architectural framework selected, we opted for a stage-based approach, which was driven by the need for stability, low cost, and low latency. Consequentially, we did not follow an agent-like approach as in previous research [32, 44]. In contrast to InteRecAgent, which employs open-ended agentic tool orchestration, EventChat limits its operation to a fixed set of five actions (Chat, Refusal, Search, Recommendation, Targeted Inquiry). This bounded controller provides predictable behavior and cost efficiency, aligning with the operational realities of SMEs. Likewise, unlike RecLLM and LLMCRS, which rely on fine-tuned dialogue models or staged subtasks, EventChat deliberately avoids fine-tuning, instead leveraging prompt-based learning to minimize the need for training data and specialist ML expertise. The need for low design complexity was primarily motivated by the startup's goal to minimize implementation costs, a requirement shared by other businesses [90]. As a result of this simplicity, the level of ML expertise required was significantly reduced compared to traditional CRS implementations not using LLMs. This is particularly important for SME adoption of LLM-driven CRS, as previous research has underscored how limited availability of AI talent inhibits AI technology adoption [5, 62, 107, 117] with technical skills identified as a specific barrier for chatbot adoption by SMEs [112]. We reason therefore that our LLM-driven CRS, with few prerequisites, minimizes the contribution of such factors, particularly as our use of prompt-based learning techniques eliminating the need for training data.

Second, EventChat integrated insights from extant RS, chatbot, and CRS research, specifically those related to conversational user interfaces (CUIs) and the incorporation of anthropomorphic features. Concerning CUI findings, research suggests that conversational systems designed to assist users in accessing information should provide search functionality and the possibility to easily access further information [54]. We supported these features, implementing the latter via a carousel cards front end component. Here, EventChat differs from prior CRS frameworks by directly linking its front end with system context: the visibility detector reports which events were actually viewed and feeds that information back into the LLM. This "visibility-to-context" loop, absent in RecLLM, InteRecAgent, and LLMCRS, grounds the dialogue in what the user has already explored, helping to avoid redundant or irrelevant suggestions. Regarding anthropomorphic features, extant findings have been mixed. While some studies have supported the use of anthropomorphic design features, for example, showing how features such as "chitchat" can improve user evaluations [54] or perceived usefulness [109], others have shown neutral [41] or even negative [115] outcomes from such phatic communication. Consequently, due to ambiguity surrounding the efficacy of anthropomorphic cues, we focused on only those features that were simple to implement and would not increase operational costs. These included the use of an avatar (a pictorial representation of the chatbot [95]), a concise self-introduction [98], and the use of ChatGPT which has been found to mimic human communication and induce anthropomorphization to a great extent [15].

Third, since EventChat was built upon the LLM ChatGPT, it inherited some of its general limitations. Among the most pressing was the tendency of EventChat to overlook information provided as context and its susceptibility to hallucinations. Additionally, there was a constraint on the context window, with a limit of 4096 tokens for input and output [86]. Specific to the system design, however, we also noticed that occasionally EventChat failed to utilize information presented in the prompt. This oversight could lead to situations, for example, where EventChat suggested events with higher prices than the expressed willingness to pay. This discrepancy could arise because the LLM would not attend to price information in the textual summary of an item during the reduction phase. For this same reason, we were able to circumvent the Refusal action in the Action Detection module when testing EventChat against prompt injection attacks (for example, see [131]). However, this only proved successful with some internal knowledge about the logic of the prompt template. Furthermore, despite our efforts to limit hallucinations through prompt design, EventChat was prone to hallucinations on rare occasions and would suggest events that did not exist in the database. This highlights another point of departure: unlike RecLLM's use of fine-tuned rankers or LLMCRS's staged subtasks, EventChat performs slate-level re-ranking of candidate items within a single prompt, which balances retrieval quality with computational efficiency. For targeted inquiries, EventChat further diverges from InteRecAgent's LLM-generated SQL approach, instead relying on digest-based QA built from structured item summaries, an approach that proved more stable under the sparse, heterogeneous event data available to the SME.

In a similar vein, there was a possibility that the answer prompt could generate responses that did not align well with EventChat's capabilities or the events available for recommendation. This problem stemmed from the answer prompt's lack of context regarding nuanced details about the startup's item corpus, such as the quality and quantity of events in the database. Consider a scenario where a user requested a highly specific type of event, for instance, "stand-up comedy". Despite EventChat finding only a few options, it might have prompted the user to refine their question for greater precision, asking users what type of stand-up comedy they preferred, which would have been an appropriate question if the user had requested a broad category like "concerts", but not for such a specific event type. The challenge lay in the fact that ChatGPT was unaware that even with a more precise description of the user's intent, the search action was likely to struggle to find a perfect match in this category. This was due to limitations in data granularity within the databases, the absence of matches, or a combination of both.

Overall, despite its simplicity, EventChat offered a range of features, including search capabilities, recommendations, and detailed information about items, thereby potentially enhancing the customer experience. These factors promote the adoption of LLM-driven CRS from a managerial perspective, successfully meeting its business objectives within various operational constraints, however with some notable limitations from a customer experience perspective due to technical trade-offs that may have negatively affected customer satisfaction with the system.

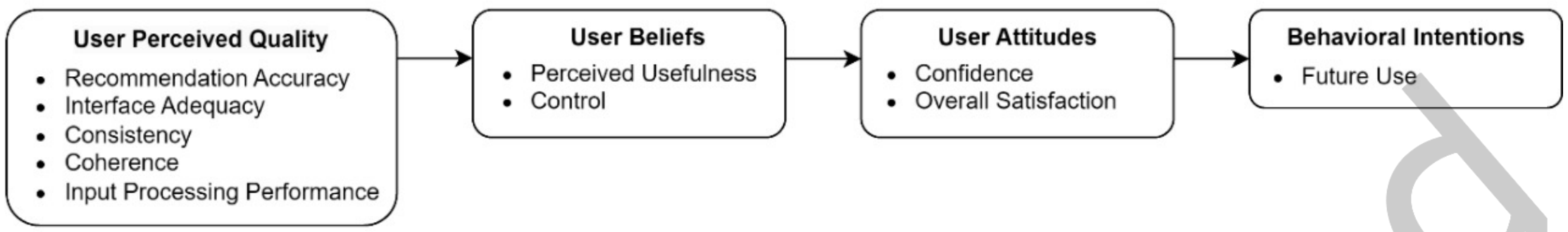

Figure 4: Conceptual CRS-adapted ResQue model

## 5 EVALUATION

To investigate our LLM-driven CRS in an SME context we conducted a field study to capture both subjective and objective evaluation metrics. This dual approach is particularly important for Gen-RecSys research, where field deployments are rare and user perceptions are not always analogous to system performance. Objective metrics included latency, token consumption (as a cost indicator), and log data concerning the system's inputs, outputs, and interim results. The subjective metrics were developed following the original ResQue model [105], further refined into a short-form version suitable for evaluation of CRS in the field, whilst also incorporating insights from Jannach's [49] catalog of subjective measurement dimensions for CRS evaluation (details in Supplementary Material 1). The following constructs were captured: Recommendation Accuracy, Interface Adequacy, Consistency, Coherence, and Input Processing Performance under the ResQue model's User Perceived Quality dimension. User Beliefs were evaluated through Control and Perceived Usefulness, User Attitudes through Confidence as well as Overall Satisfaction, and Behavioral Intentions through Future Use (Figure 4). Notably, we introduced the constructs of Consistency, Coherence, and Input Processing Performance, expanding the traditional Perceived Quality dimension to effectively assess key conversational quality criteria of CRS. Single-item measures were used for all constructs to minimize participant burden in a real-world field setting, in line with previous applications of ResQue in applied domains. While this reduces internal redundancy, it was essential for ensuring survey completion and external validity. Our dual approach of incorporating both subjective and objective metrics allowed us to gain insights into both strategically important metrics such as user evaluations, costs, and load times, as well as synthesize the findings in light of the technical implementation via log data.

### 5.1 Methodology

#### 5.1.1 *Survey design*

We conducted a field study with real users updating or downloading the EventChat app. For this reason, we minimized participant burden by including only one item per construct, in line with previous applications of ResQue model to applied contexts [26, 63]. The survey was designed to assess the overall quality and user experience of the CRS, rather than examining its individual components. This choice was in accordance with recommendations for applying ResQue [106] and based on the understanding that evaluating the components would primarily reflect the performance of the underlying technologies, such as the LLM ChatGPT or the RS Amazon Personalize, rather than the overall user experience of EventChat. All items were assessed using a 5-point Likert scale ranging from 1 (disagree) to 5 (agree), to ensure that all items were displayed correctly across a variety of smartphone devices. In addition to the ResQue items, we also asked respondents if their request was successfully fulfilled (Success). For successes, we asked for Perceived Effort as per Loepp et al. [80]. Independent of success, we asked for open-ended feedback on problems the users experienced in using the CRS (General Problems).

#### *5.1.2 Objective User-interaction Metrics*

For each user interaction with EventChat, we monitored several key metrics: latency, token usage, and log data. These metrics were recorded for every request made to the Microsoft Azure OpenAI API, corresponding to each prompt call. Additionally, we measured the total duration of a request on the application's front end, including the round-trip time to the server. Latency metrics served as an estimate of the user's loading time, while token usage data directly provided insight into operational costs, both of which are vital strategic considerations for businesses. By analyzing these metrics at the level of individual prompts, we were able to pinpoint the most significant performance bottlenecks. Furthermore, the collection of log data enabled us to analyze ChatGPT's output and identify irregularities as well as edge cases. These insights allow for future improvements of the system.

#### *5.1.3 Data Collection*

Participants accessed the survey via a link labeled as 'Lottery' in the app (Figure 2) that offered participants a chance to win one of three Amazon vouchers worth €50 each. Data collection occurred between 18.10.2023 and 14.12.2023. Respondents were either: (i) existing users of the startup's app who accessed the corresponding screen of EventChat, or (ii) individuals recruited on Instagram from 29.10.2023 to 11.11.2023 using a total campaign budget of €22 (24 USD). Our sample thus consisted of both existing and newly recruited users, thereby capturing a broader cross-section of user types, improving external validity. A total of 108 participants conducted the survey. The university ethics board reviewed and approved the study prior to data collection, proposal no. 2023-N-247. After filtering out non-completes (n = 20) and those who did not interact with EventChat according to the log data (n = 5), a total of 83 observations were available for analyses (43.8% female; Mage = 28.3, SD = 11.0). Only the log data, cost, and latency for these 83 survey participants that used the app were analyzed to ensure the completeness of the usability metrics and survey responses.

### 5.2 Results

#### *5.2.1 Subjective Ratings of the User Experience*

To test subjective measures of the user experience, we analyzed our short-form ResQue CRS model using a path analysis (i.e., structural equation model using observed variables only). Results showed that 85.5% (n = 71) of users self-reported Recommendation Accuracy as neutral or good, showing respectable performance for a newly developed system in the field. Recommendation Accuracy, however, also simultaneously resulted in the highest number of negative appraisals of all User Perceived Quality and User Beliefs constructs (15%, n = 12), with a further 66.7% (n = 8) of these users indicating that EventChat failed to fulfill their request(s) (Success), contrasting the overall Success rating of 83.1% (n = 69). The findings thus underscore the importance of refining the core components after an initial evaluation, indicating that improvements to recommendations would be amongst the most pressing changes in future systems. Descriptive statistics, visualization of the survey results, and quotes regarding reasons for app failure can be found in the supplementary material 2.1, 2.2, and 2.3.

Figure 5 shows the responses for Perceived Effort. Analysis of the corresponding log data revealed a median of 2 turns to identify a suitable event, consistent with the low perceived effort reported by most participants. However, three respondents indicated significant loading times in the General Problems section, which was subsequently confirmed by the high latency metrics detailed in section 5.2.4. Notably, among these, two rated the effort as high, while one reported it as low. This variation suggests a divergence in user expectations or tolerance concerning response times. It appears that the efficiency in locating suitable events may have offset the impact of loading times for most users.

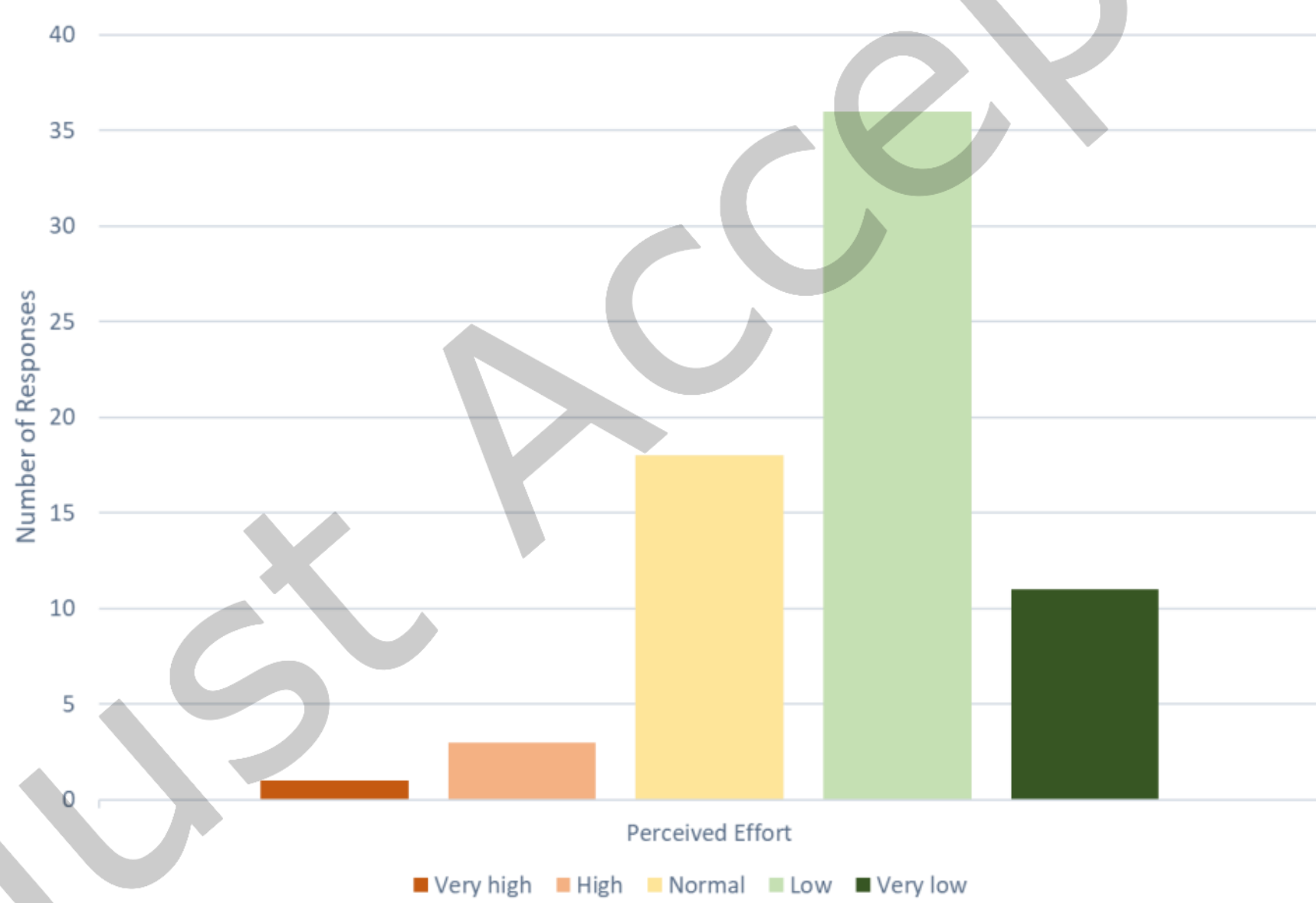

Figure 5: Perceived Effort of participants to find suitable events

#### 5.2.2 *Root Cause Analysis: Understanding why EventChat failed to fulfill requests*

To investigate the underlying reasons why EventChat occasionally failed to fulfill users' requests (Success), we examined the logs of the corresponding sessions. We identified the following main failure categories: missing relevance of suggested events, failed Targeted Inquiry actions, and issues defining the time and location of the suggested events. In the following passages, we elaborate on each aspect and synthesize the evaluation with technical limitations:

The most reported issues concerned the relevance of suggested events. In many cases, the requested events did not exist in the database or were not properly categorized (e.g., category "Other") as a consequence of the startup's web-scraping strategy. Other problems can be attributed to our simple design of the Recommendation action. While we extracted query parameters including the event category for the Search action, which we either included in the request to the startup's Amazon Personalize or its vector store, we did not apply filters beyond the time interval for the Recommendation action. We adopted this approach from Wang et al. [123] who reported competitive performance. Consequently, we can attribute this underperformance to the creation of the candidate set by calling the startup's Amazon Personalize instance without incorporating the expressed preferences of the user via additional filters.

Furthermore, several participants encountered issues with EventChat in recognizing their questions related to an event (Targeted Inquiry action). In most cases EventChat either opted for the Refusal action or failed to establish a connection between the question and one of the events stated in the prompt as part of the chat history. This highlights challenges in providing comprehensive information solely through prompt-based learning techniques. We also found other limitations of prompt-based learning beyond this root cause analysis. Particularly, respondents indicated a language switch from English to German in their conversation, which we confirmed through log data. In these cases, ChatGPT would not adhere to the instructions defining the response language as English but would be biased by German event descriptions which made up the majority of textual information within the prompt.

Finally, despite users stating that the interface was intuitive to use, in many cases, users did not seem to use the time button in the interface but rather communicated their time preferences using the chat. In these scenarios, the candidate set was generated without the application of time filters. Consequently, during the reranking phase, ChatGPT would not consider the user's specified time information, resulting in the suggestion of events that do not align with the user's time requirements. Overall, a more user-friendly CRS would, therefore, extract all the necessary information from the chat.

These observations highlight the practical trade-offs of a prompt-based SME deployment: while sufficient for basic interaction with users, context handling and personalization remain improvable without additional data or fine-tuning.

#### *5.2.3 Results from Path Analysis*

The model showed adequate fit across most indices ($\chi^2$ (22) = 33.942, p = .05, CFI = 0.948, TLI = 0.917, RMSEA = 0.081, SRMR = 0.087). Reporting path coefficients between User Perceived Quality and User Beliefs constructs first, results showed that self-reported Recommendation Accuracy ($\beta$ = .36, $p < .01$) and Consistency ($\beta$ = .352, $p < .01$) were positively associated with higher ratings of Perceived Usefulness. Input Processing Performance was positively associated with higher Control ($\beta$ = .319, $p < .05$), and Control in turn was positively associated with higher ratings of Perceived Usefulness ($\beta$ =

.202, p < .05). Examining the paths between User Beliefs and User Attitudes constructs showed that Perceived Usefulness was significantly linked to higher Confidence (β = .683, p < .001) and Overall Satisfaction (β = .513, p < .001). Lastly, examining the relationship between User Attitudes and Behavioral Intentions, we found that Overall Satisfaction significantly linked with higher Future Use (β = .503, p < .001).

In sum, the path analysis confirms that constructs introduced for conversational quality (e.g., Consistency, Input Processing Performance) are indeed predictive of user beliefs and attitudes, validating their inclusion in the revised ResQue model for LLM-driven CRS. A full table containing path coefficients can be found in supplementary material 2.4 and a visualization is presented in Figure 6. In supplementary material 2.5, we present additional results that explored how the model could be improved using the model's modification indices, that may be informative for other researchers wishing to extend or revise this model for other Gen-RecSys deployments. In short: results show that model fit remained stable across specifications, with gradual improvements as weakly contributing relationships were removed, thus indicating that the revised ResQue model captures Gen-RecSys evaluations in both a parsimonious and robust manner.

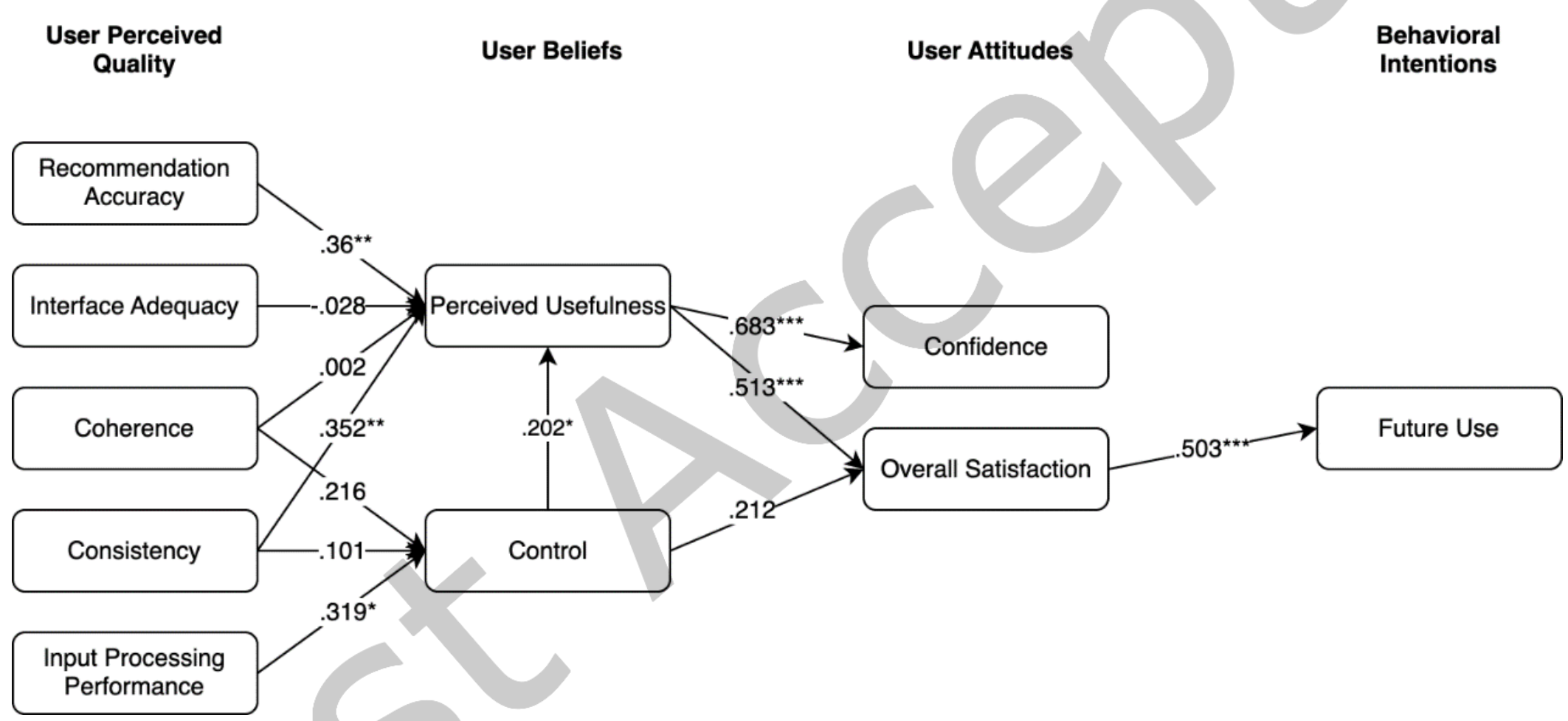

*Note:* *: p < .05, **: p < .01, ***: p < .001

Figure 6: Path analysis results.

### *5.2.4 Performance Metrics*

Next, we examine the token usage and latency in EventChat. Table 2 presents these metrics, aggregated over all chat sessions of the survey participants. With a median latency of 5.7s per message and high token usage, EventChat highlights the cost-latency trade-offs central to SME adoption. Most critically, the reduction stage consumed disproportionate resources, underscoring that LLM-based re-ranking is a key bottleneck for practical deployment. The reported interaction costs reflect metered token usage from ChatGPT calls recorded during live operation. They therefore represent observed operational token expenditure per interaction rather than a simulated cost model.

Table 2: Token usage and latency metrics for chat sessions associated with survey participants.

| Median tokens used per chat message | Median tokens used per chat session | Median latency per message | Median latency per chat session |
|---|---|---|---|
| 18106 | 56325 | 5.7s | 13.7s |

Table 3 displays the consumption of tokens as well as the needed time in more detail per action or stage. The data shows that the reduction of candidate items, as derived from the Search or Recommender prompts, consumed the most resources.

Table 3: Token usage and latency per module or phase.

| Phase/ action | Median tokens used per chat message | Median latency per message |
|---|---|---|
| Action Detection (including Chat, Refusal) | 2622 | 2.7s |
| Targeted Inquiry | 852 | 0.6s |
| Search | 1724 | 1.6s |
| Recommender | 796 | 1.2s |
| Reduction | 23408 | 4.0s |
| Answer creation | 2419 | 2.6s |

*Note:* Median of the sum of tokens used per Reduction module call

Utilizing the median token per phase displayed in Table 2 and 3, a medium cost of $0.04 per interaction can therefore be inferred based on the Azure OpenAI per-token pricing in August 2023. This estimate reflects LLM token usage only and excludes supporting infrastructure (e.g., database, hosting services etc.) within the startup's existing environment. As costs scale linearly with token consumption, the presented figures can be used to approximate costs under alternative pricing schemes. Supplementary material 2.6 and 2.7 contain contains an extended analysis of the performance metrics, incorporating all interactions regardless of their survey completion status to better account for potential selection bias in these results. Results were, however, highly comparably to those presented in Table 2 and Table 3.

## 6 DISCUSSION

Implementing a CRS has become simpler than ever before thanks to recent advances in LLMs, creating new opportunities for SMEs to offer innovative recommendation features that enhance customer value beyond traditional approaches. Yet, in this emerging field, numerous open questions remain for businesses considering Gen-RecSys adoption. Throughout this paper, we have discussed our design choices and examined their impact on business-critical factors. Our LLM-driven CRS, EventChat, adopts a stage-based architecture for stability and relies on prompt-based learning to reduce complexity. Results from the field show that the system effectively delivers core CRS functions, namely successful recommendations and a satisfactory user experience, and could be deployed as a minimal viable product with real customers. At the same time, long-term success would require addressing issues around cost, speed, and recommendation quality. Incremental improvements may include fine-tuning the LLM through reinforcement learning from human feedback [32, 99] to better align outputs with user needs, enabling SMEs to adopt an agile development strategy that systematically refines system performance. In parallel, integration with newer foundation models will need continual reassessment, given the frequent release of more powerful alternatives that competitors may adopt.

## 6.1 System Design Implications and Transferable Insights

Building on the principles outlined in previous research and the lessons learned from working with the company and performing the empirical evaluation, we have identified five design principles that aim to enhance LLM-driven CRS for SMEs, addressing the significant challenges and opportunities for their design.

### *6.1.1 Design Insight 1: Stage-based architectures can be beneficial relative to agentic pipelines in SME contexts.*

During the implementation process, we encountered a critical trade-off between quality of the user experience and constraints of cost and latency, which was particularly evident in the architecture and prompt design. Architecturally, we opted for a stage-based approach, contrasting previous research that has generally emphasized the potential of agent-based approaches (e.g., for industrial automation [130] or chemical engineering [13]). While we experimented with an agent-based approach, preliminary tests showed that agent-based orchestration resulted in increased LLM call frequency and inconsistent tool invocation behavior, leading to higher latency variance. Combined with the measured token cost dominance of the reduction stage, this motivated our adoption of a stage-based architecture which addressed stability issues with ChatGPT and aligned with the business requirements of an SME regarding latency and costs. At the prompt design level, this contrasts research that focuses on optimizing LLM performance through various prompting strategies [19, 66, 87, 135] while often overlooking business-critical trade-offs. Notable exceptions exist in the broader LLM literature, such as Zong et al. [136] on cost-quality balances and Hämäläinen et al. [40] on data quality versus latency in synthetic data generation. Our contribution adds to this stream by providing an empirical perspective from a real SME deployment, showing how these trade-offs manifest in practice. More broadly, our experience suggests that when cost predictability and latency are primary concerns (as is often the case in SMEs) stage-based orchestration may offer greater operational stability than more flexible but potentially expensive agentic pipelines. This aligns with broader discussions in the LLM-agent literature that emphasize the added complexity and evaluation challenges introduced by multi-step tool orchestration in agentic systems [126].

#### 6.1.2 Design Insight 2: Prompt-only approaches face quality limitations as contextual complexity increases.

We found that solely relying on prompt-based learning poses quality issues. A key factor driving low user evaluations was ChatGPT's occasional oversight or misinterpretation of contextual information, as revealed by our log data. Our approach, inspired by the InteRecAgent framework [44], which relied heavily on prompt-based learning techniques like ICL or Few-shot CoT without model fine-tuning. While these techniques offer simplicity and flexibility well-suited for SMEs, our use case required longer prompts to capture detailed information. This content was sometimes missed or misunderstood by ChatGPT, consistent with findings in recent research [132], and led to lower Consistency and Coherence ratings. Similarly, it proved unfeasible to design a single comprehensive prompt covering all eventualities due to context window limits, cost, and latency. This limitation would persist even with agent-based approaches. While a unified dialog manager, as proposed by Friedman et al. [32] could improve responses by recognizing its tools and avoiding unfeasible suggestions, this would not resolve the issue of missing detailed contextual knowledge. Our findings therefore underscore the limits of prompt-only approaches and point to the need for answer generation that incorporates finer-grained corpus details. This suggests that prompt-only architectures may be sufficient for relatively simple or well-bounded recommendation tasks, but may require augmentation when domain knowledge, personalization depth, or contextual nuance increases.

#### 6.1.3 Design Insight 3: Production deployments require guardrails to mitigate hallucination and prompt-injection risks.

Our deployment revealed that LLM-driven CRS are susceptible to occasional hallucinations and prompt-injection vulnerabilities, the former being a well-known behavior of LLMs and the latter widely recognized as a security risk in LLM applications [28]. In rare cases, EventChat suggested items that did not exist in the database or failed to adhere strictly to contextual constraints. While we employed schema-based output formatting and a bounded five-action controller to limit uncontrolled generation, these mechanisms do not eliminate all risks. For SME deployments in particular, where trust and reliability are critical, additional architectural guardrails may be advisable. These include constraining outputs to validated item identifiers, verifying generated content against the underlying database prior to rendering recommendations, and carefully isolating tool-calling logic from free-form user prompts. Although not all such mechanisms were implemented in EventChat, our experience highlights their importance for production-grade LLM-driven CRS.

#### 6.1.4 Design Insight 4: Hybrid chat interfaces should align with natural conversational behavior.

Our analysis suggests that all relevant parameters for recommendation should be directly communicated throughout the conversation. Users frequently entered time preferences via the chat interface rather than using the dedicated button. This indicates that, despite their potential to simplify system design, static buttons may not align with intuitive user behavior when placed outside the conversational flow. This represents a limitation of hybrid chat-button interfaces in LLM-driven CRS, even though such approaches have previously been linked to higher perceived control [47, 92]. More generally, interface mechanisms intended to reduce ambiguity or iteration costs should be evaluated against observed user behavior, as design assumptions about control and efficiency may not be reflected in actual usage patterns.

#### 6.1.5 Design Insight 5: LLM-based reranking can become economically prohibitive at production scale.

We found that using a several-billion-parameter LLM as a re-ranker in the RAG technique is prohibitively expensive. This reflects a broader pattern documented in LLM deployments, where model selection and architecture depend on explicit cost–quality trade-offs [136]. The reduction phase was a major cost driver with a median of 4.6 cents per message, given current API pricing [86]. With a median of 4s, it also significantly contributed to latency. Although latency can be reduced by adopting an item-level reranking approach, as done in RecLLM [32], this further increases costs. Consequently, our results suggest that approaches relying on large LLMs for reranking [32, 125, 133] are economically unviable for most production settings, even when using smaller LLM variants. For SMEs in particular, where margins and traffic volumes directly amplify API expenditures, LLM-enhanced components should be evaluated not only in terms of recommendation alignment but also with explicit cost–latency accounting.

### 6.2 Managerial Implications

Our case study reveals key managerial implications for SMEs when adopting LLM-driven CRS. First, the financial aspect of deployment is a significant factor. In our implementation, operational costs averaged 3.6 cents per message and 11.2 cents per chat session implying that such systems are currently most viable in high-margin scenarios. Latency is another critical factor: with an average of 5.7 seconds per message and 13.7 seconds per session, managers must weigh the value of improved customer experience against potential wait times. Encouragingly, our user survey indicated low perceived effort, suggesting that efficiency gains may offset latency concerns for many users. Still, both costs and latency should be seen as recurring challenges for any system dependent on external LLM APIs.

Second, the complexity of long-term production use should not be underestimated. Although libraries such as Langchain [119] appear to lower entry barriers, our experience shows that implementing a LLM-driven CRS that is robust over time remains demanding. On the one hand, reliance on prompt-based learning reduces the need for advanced ML skills and thus lowers adoption barriers for SMEs [4, 111, 112]. On the other hand, this approach introduces limitations in quality, while alternatives, such as adopting more advanced LLMs or pursuing fine-tuning, introduce their own cost and complexity trade-offs [43]. Moreover, even if not observed in our field study, managers must anticipate the additional need for safeguards against abusive usage when operating customer-facing systems.

Taken together, our findings suggest that LLM-driven CRS can be feasible for SMEs but only when implemented with careful cost-benefit planning. This echoes Ivanov and Webster's [48] call for rigorous evaluation of AI adoption and underlines the "no free lunch" principle [2]: every technical choice entails trade-offs that must be aligned with the firm's objectives, margins, and operational capacity.

### 6.3 Theoretical Implications

This paper contributes to theory by exploring a novel method for the user-centric evaluation of LLM-driven CRS. We are the first to apply an adapted short-form ResQue model to the evaluation of an LLM-driven CRS, launched in the field with real users. We therefore lay the foundation for future studies to expand its usage in LLM-driven CRS settings. Such an approach is vital for deepening our understanding of key user issues with the technology and the motivations behind its acceptance [106], as well as for tracking advancements within the domain [49]. The significance of our research is also underscored by the growing interest in CRS in both academic [82] and practical fields [34]. More critically, we address the rising interest in LLM-driven artifacts, a technology that has been identified as being set to revolutionize digital services [30].

Our results, in part, reinforce the continued relevance of relationships established in earlier frameworks. Specifically, Perceived Usefulness was positively associated with user satisfaction and intention to use, self-reported Recommendation Accuracy with higher Perceived Usefulness, and User Control with higher overall user Satisfaction. These findings align with extant research in RS [6, 16, 42, 53, 59, 60], conversational agent domains [108], and the original ResQue framework [105], showing how elements of the original ResQue model remain relevant. This is important contribution, as it suggests that core models and constructs for evaluating technologies of the past transfer may in part transfer to the era of generative AI, helping to shape information systems research in the coming years. Nevertheless, our findings also reveal important deviations from existing work. Within the User Perceived Quality dimension, Coherence and Interface Adequacy were insignificant predictors, contrasting Jannach's [48] recently proposed catalog of CRS evaluation dimensions, which had not to date been empirically verified. Our findings also validated the inclusion of Input Processing Performance and Consistency, constructs recently proposed in human-AI interaction research [18, 23]. Consequentially, we interpret these mixed findings as theoretically informative: The non-significance of Coherence and Interface Adequacy suggests that certain conversational quality dimensions may be context-dependent, particularly in short, task-oriented interactions in a leisure context where recommendation relevance and immediate responsiveness likely dominate user evaluations. In other contexts, where well-reasoned recommendations are vital for human-AI collaboration (e.g., legal advice, education), we posit that Coherence could still have predictive utility.

Taken together, these findings underscore the importance of empirically validating evaluation dimensions as CRS architectures evolve, rather than assuming that either legacy or newly proposed constructs will generalize across contexts. With this in mind, and compared to alternative proposals that advocate adding many new constructs [49] our results suggest a more parsimonious model for Gen-RecSys evaluation may suffice. This emphasis on parsimony is reinforced by our design choice to use single-item measures, which demonstrated adequate model fit, and supports the viability of a short-form ResQue for field studies. Theoretically, we therefore contribute a validated, concise evaluation framework tailored to LLM-driven CRS, offering a roadmap for revising established acceptance models for new AI applications.

Finally, our study illustrates the benefits of combining subjective evaluations with objective system metrics. Log analysis revealed how prompt-based learning produced inconsistencies and hallucinations, affecting both conversational and recommendation quality. Users valued consistency, as it reduced the need to correct system errors, an intuition noted in prior conversational agent research [56, 77] but not previously demonstrated empirically. By triangulating subjective and objective measures, we highlight a broader theoretical issue: the blurred boundary between evaluating the LLM itself and the surrounding CRS framework. While LLM performance is central to system effectiveness, disentangling improvements in models from improvements in system design remains a challenge, underscoring the need for evaluation frameworks that explicitly account for this interdependence.

In sum, our contributions demonstrate how existing frameworks may be sufficient and where changes may be necessary: while established constructs retain predictive power, new dimensions are needed to capture conversational dynamics, and parsimonious models can be effective in practice. Our work therefore advances the development of user-centric evaluation frameworks for LLM-driven CRS and responds to recent calls for criteria that align evaluation with user expectations rather than solely technical performance [124].

### 6.4 Limitations

This research has several limitations that should be acknowledged. First, the CRS implementation and data presented were contextualized to the needs of a startup in the leisure industry. While this single context may limit generalizability, our use case nonetheless exhibited characteristics that are common to a variety of SMEs, including a diverse set of item categories, sensitivity to user pricing, and time dependence in recommendations. These characteristics add realistic complexity, but also mean that certain design principles highlighted may not directly transfer to larger firms that face fewer resource constraints.

Second, our study employed SEM with a modest sample size, and while the SME's customers are known to be young due to the leisure-oriented context, extensive user background data were not collected in the current sample due to the administrative and legal challenges for small firms operating within the EU (i.e., compliance with strict GDPR regulations). Consequentially, we are unable to control for possible selection bias that may affect the generalizability of our results. Additionally, the use of single-item measures for our constructs may reduce internal validity. We aimed to mitigate this by using well-established measures from the RS domain [49, 105] and the straightforward nature of the constructs under investigation [9]. On the other hand, the strength of our data lies in its external validity, as the data represent interactions from actual users of an LLM-driven CRS in the field. Our work thus offers real-world insights that controlled environments face difficulty replicating. Third, to reduce participant burden, we refined the ResQue model to shorten the number of collected constructs while adhering to the CRS principles outlined by Jannach [49]. In line with the recommendations of Jin et al. [58], we excluded constructs that were less relevant for our empirical context or system design. For instance, Recommendation Novelty was excluded because only upcoming events were recommended, and Information Sufficiency was excluded because users could access additional details by clicking on cards. While these omissions were both necessary and informed by theory, they may nevertheless have predictive power in other CRS contexts.

Third, we did not perform ablation studies to isolate the effects of different LLMs and/or components, which would be important to demonstrate the degree of robustness of our system design on system performance. Our implementation used ChatGPT, which was the most advanced LLM on Azure OpenAI Service at the time. The subsequent release of larger and more efficient LLMs (e.g., GPT-4) and those incorporating greater reasoning capabilities (e.g., Gemini 2.5 Pro) may mitigate weaknesses relating to overlooking context or hallucinations via their extended context windows and improved factual grounding, albeit at a likely increase in costs. Therefore, our discussion on trade-offs faced by SMEs remains relevant. EventChat's design choices (fixed action controller, slate-level re-ranking, digest-based inquiry) illustrate one valid route for bounding costs while enabling conversational recommendation for SMEs. Future research with industry partners may, however, use ablation studies to quantify the trade-off between advanced LLMs and system components on costs, quality, and latency.

Finally, this study was designed as a field deployment and evaluation under real operational constraints rather than as a controlled benchmarking study of recommendation algorithms or system architectures. As such, our findings should be interpreted as deployment-based evidence documenting feasibility, measurement practices, and design trade-offs within a live SME implementation. Future work may build upon this deployment report through controlled comparisons and multi-site replications to further assess robustness and transferability.

### 6.5 Future Research

In elaborating the use of LLM-driven CRS, we recommend the following technical, strategic and theoretical research directions (RD):

One practical avenue would be exploring use of an LLM featuring more parameters like GPT-4o (RD1) that could allow for agent-based approaches [44]. (RD1). An alternative path is to investigate the use of smaller fine-tuned LLMs, as in Friedman et al. [32] and Feng et al. [29], which may better control these trade-offs by reducing both computational demands and operational expenses (RD2). Beyond model choice, future research should also consider resource-efficient methods for generating training data and quantifying the costs associated with fine-tuning and deployment (RD3). This direction is particularly relevant in SME contexts, where access to large text corpora for training is often limited. In parallel, future frameworks for LLM-driven CRS should examine strategies for incorporating fine-grained and often subtle domain knowledge into answer generation (RD4), as our findings highlighted the limitations of relying on prompt-based learning alone. Finally, there is a need for systematic evaluations of different model configurations within a given system design. Structured ablation studies comparing larger general-purpose LLMs, smaller fine-tuned models, and hybrid approaches would help establish the balance between component design and foundation model choice, clarifying how best to achieve low cost, low latency, and high quality in deployment (RD5)

From a managerial perspective, our work has identified several factors to consider when evaluating the value importance of adopting a CRS. Nonetheless, as Jannach [49] observes, a significant gap remains regarding the broader determinants of CRS adoption from a business perspective, particularly the relative benefits of conversational recommendation compared to traditional search methods (e.g., filters). Prior research suggests that user satisfaction can sometimes be higher with menu-based interfaces than with chat-based interfaces [92]. Yet our findings showed that users often ignored buttons integrated into the interface and instead communicated their demands directly in the chat. Thus, further research should investigate how and when interface design can complement the chat (RD6). Another direction concerns the firm-level drivers of adoption. Future work should investigate whether factors such as existing IT infrastructure, which may enable rapid integration of conversational AI, or unique strategic needs for improving customer-company information exchange, act as important motivators for firms (RD7). Such studies would complement the current focus of RS research, which tends to concentrate on end users [76], by highlighting the organizational conditions that shape adoption decisions. Exploring these aspects could involve empirically assessing the business value attributed to LLM-driven CRS as previously explored in RS domains [101] or developing a decision-making framework similar to that taken by Schuetzler et al. [111] in the chatbot domain, for example.

Turning to theoretical considerations, while recent research has adapted the ResQue framework for the evaluation of CRS [58], ambiguity remains on how to evaluate LLM-driven CRS, which will likely dominate future RS applications. Building on our findings, we argue for developing a unifying user-centric evaluation framework for LLM-driven CRS and Gen-RecSys more widely (RD8). Future research could examine revisions to the short-form evaluation model we presented. Our expanded analyses (see supplementary material 2.4) showed that eliminating Coherence and Interface Adequacy improved model fit. Future research could seek to confirm these findings with larger sample sizes (RD9). Moreover, applying new methods to evaluate LLM-driven CRS is a promising direction (RD10). Streamlined methods that remain relevant for assessing overall system quality could leverage simulations of user interactions using LLMs, drawing on innovative concepts such as using LLMs for assessing outputs from LLMs [33, 126] or the user simulation-based CRS evaluation framework developed by Zhang and Balog [134].

## 7 CONCLUSION

Our case study demonstrates the technical feasibility of implementing and deploying a large language model (LLM)-driven conversational recommender system (CRS) for small and mid-sized enterprises (SMEs). We introduced a short-form model for LLM-driven CRS evaluation suited for field studies, providing a framework for future research. At the same time, our evaluation revealed challenges related to latency, operational costs, and the quality of user interactions, which form a barrier to long-term usage. We identified several design choices giving rise to these issues, including our reliance on prompt-based learning and the use of ChatGPT as a ranker in the retrieval-augmented generation (RAG) technique. This underscores the need for innovative approaches for the architecture of LLM-driven applications.

Our findings highlight that the successful implementation of LLM-driven CRS by SMEs depends on an in-balancing several interrelated factors: acceptable user waiting times, the profitability of use cases, and realistic feature expectations, all of which influence system complexity and costs. By grounding these insights in the real-world SME context provided by *EventChat*, our work extends the Gen-RecSys literature beyond algorithmic innovation to emphasize the organizational and managerial conditions that will determine whether LLM-driven CRS can achieve widespread adoption.

## NOTE

Please note that the following article is an expanded version of a pre-print shared by the authors on arXiv [71]. To support reproducibility, partial code (frontend and backend modules, survey instruments, and path analysis scripts) have been uploaded to a public repository [70]. A reproducibility checklist is included in supplementary material 2.8.